\documentclass[a4paper,11pt]{article}
\usepackage{epsfig,epic}
\usepackage{wrapfig}

\newcommand{\beq}{\begin{equation}}
\newcommand{\eeq}{\end{equation}}
\newcommand{\eqa}{\begin{eqnarray}}
\newcommand{\ena}{\end{eqnarray}}
\newcommand{\nonu}{\nonumber}
\begin{document}
\begin{titlepage}
\begin{flushright}
SISSA 50/2003/FM\\
GEF-TH-06/2003\\
\end{flushright}
\begin{center}
{\LARGE \bf On the magnetic perturbation \\
 of the Ising model on the sphere}
\end{center}
\vskip 0.8cm
\centerline{
P. Grinza$^a$\footnote{e--mail: grinza@sissa.it} and
N. Magnoli$^b$\footnote{e--mail: magnoli@ge.infn.it}}
 \vskip 0.6cm
 \centerline{\sl  $^a$ SISSA and INFN}
 \centerline{\sl via Beirut 2-4, I-34014 Trieste, Italy}
 \vskip .2 cm
 \centerline{\sl  $^b$ Dipartimento di Fisica,
 Universit\`a di Genova and}
 \centerline{\sl Istituto Nazionale di Fisica Nucleare, Sezione di Genova}
 \centerline{\sl via Dodecaneso 33, I-16146 Genova, Italy}
 \vskip 0.6cm

\vspace{1.0cm}

\begin{abstract}
In this letter we will extend the analysis given by Al. Zamolodchikov 
for the scaling Yang-Lee model on the sphere to the Ising model in 
a magnetic field.
A numerical study of the partition function and of the vacuum expectation
values (VEV) is done by using the truncated conformal space (TCS) approach.
Our results strongly suggest that the partition function is an entire function
of the coupling constant.  
\end{abstract}
\end{titlepage} 
 
\setcounter{footnote}{0} 
\def\thefootnote{\arabic{footnote}} 
 
\section{Introduction} 

Quantum field theories on curved (fixed) background have attracted much attention and have been explored for a long time \cite{birrel}.
The main reason to study them is their importance as a first step toward the understanding  of quantum gravity. 
In this letter we will consider 2D 
conformal field theories (CFT) and their perturbations on a spherical background.
Apart from their relevance in order to understand 2D quantum 
gravity, the theories on the sphere can be considered as infrared 
regularizations of the corresponding theories defined on the infinite plane;
these latter can in general be recovered in the limit of infinite radius.
This natural cutoff avoids the presence of infrared divergences in the perturbative theory and the various physical 
quantities are analytic in the coupling constant $\lambda$ at $\lambda =0$ (see also the footnote in sect. 4). 

In a recent paper Al. Zamolodchikov \cite{zamo} proposed a novel way to study the 
partition function of the scaling Yang-Lee model  
on the sphere. 
Motivated by a numerical analysis he suggested that such partition function is an entire function of the coupling constant. 
Assuming that this conjecture is true,
one can get useful informations about the large $R$ limit ($R$ being the radius of the sphere) 
from the knowledge of the first few terms of the expansion coming from Conformal Perturbation Theory. \\
It is interesting to test this approach considering other relevant perturbations of Minimal Models, 
in particular we will study the Ising model 
perturbed by a magnetic field. The former conjecture has been confirmed analytically \cite{zamo} 
in the case of the thermal perturbation of the Ising model resorting upon the fact that it 
is equivalent to a free massive Majorana fermion. The present case is less 
trivial (it is not a free theory) and requires to be investigated with the same numerical methods 
employed in \cite{zamo}. In particular we addressed two main issues: the asymptotic behaviour 
of the Vacuum Expectation Values (VEV) on the sphere; the asymptotic location of the zeroes of the partition function.  \\ 
This  letter is organized in the following way: 
In sec.~ 2  we will briefly introduce CFT on the sphere and the perturbed theory, 
in sec.~  3  we will describe  the numerical results and finally in sect.~  4 we will give our  conclusions. 
 
\section{Ising model in a magnetic field} 
In the recent past, Conformal Field Theories (and in particular Minimal Models) have been  
studied on a general Riemann surface \cite{eguchi}. In the following we will deal with models defined on a sphere,
\footnote{We assume that there are no conical singularities and the metric is smooth. In the presence of conical singularities
there are metric dependent terms in the partition function \cite{cardy} .} 
the simplest non-trivial example of curved geometry. 
In particular,  we will consider the first model of the minimal unitary series, i.e. the Ising model. 
It is characterized by a set of primary fields
$\phi_i$ ($1$, $\sigma $ and $\epsilon$) 
which transform in the following way 
\begin{equation} 
\delta \phi _i (x) = -\Delta _i \phi _i (x) \delta \chi (x)  ~~~~~~~~ g_{ab}=~e^\chi \delta _{ab}
\end{equation} 
under a Weyl transformation 
\begin{equation} 
\delta g_{ab} (x) = g_{ab}(x) \delta \chi (x) 
\end{equation}      
where 
\beq
e ^{\chi (x)} = \frac{4 R^2}{1+z \bar{z}}
\eeq
is the Weil  factor of the metric,
$\Delta _i $ are the conformal weights of primary fields,  
given respectively by $0,1/16$ and $1/2$. 
 
The trace of energy-momentum tensor in CFTs defined upon non-trivial geometric backgrounds plays a central r\^ole. In fact, it gives a quantitative characterization of the effect of a change in the geometry. Its explicit form for the case of the sphere (with radius $R$) is given by
\begin{equation} 
\label{confano} 
\theta (x) = -\frac{c}{12} \hat{R} 
\end{equation}     
where $ \hat{R}$=$\frac{2}{R^2}$  is the scalar curvature of the sphere and $c$ =$1/2$ is the central charge. 
 
\noindent 
By eq. (\ref{confano}) and by the definition of  
$\theta (x)$ in terms of the partition function $Z_{CFT} (R)$ 
one gets  the relation 
\beq 
Z_{CFT}(R) = R^{c/3} Z_0 
\eeq 
where $Z_0=Z_{CFT}(1)$. 

\vspace{0.5cm}
As shown in \cite{zamo}, the action of a CFT perturbed by a relevant operator is given by   
\beq 
\mathcal{S}_{\lambda}=\mathcal{S}_{CFT}+\frac\lambda{2\pi}\int 
\sigma(x)e^{\chi(x)}d^{2}x\label{cpert} 
\eeq 
$\mathcal{S}_{CFT}$ is the action of the Ising model on the sphere and $\sigma(x)$ is the perturbing operator.  
The partition function $Z_ \lambda (R) $ can be calculated in the regimes of both small and large values of $R$. 
 
\noindent 
In the first case, 
expanding in $\lambda$ and defining  
\beq 
h=\lambda (2R)^{2-2\Delta_\sigma} 
\eeq 
and 
\beq 
z(h)=\frac{Z_{\lambda}(R)}{Z_0 R^{c/3}} 
\eeq 
one gets 
\beq 
z(h)=\sum_{n=0}^{\infty}(-h)^{n}z_{n}\label{zn} 
\eeq 
where  
$z_0 =1$ and  
\beq 
z_n =\frac{\pi } {(2 \pi )^n n!} \int \langle \sigma (0) ... \sigma(y_n ) \rangle \prod_{i=2}^{n}  
\frac {d^2 y_i} {(1 + y_i \bar{y}_i)^{2-2 \Delta_\sigma}}. 
\eeq 
and  the correlators are calculated in the conformal theory on the plane.
By using the fusion rules of Ising model it is easy to show that only  even  coefficients 
are different from zero. 
 
\noindent 
In \cite{zamo}, it is conjectured that this series is absolutely convergent 
and defines an entire function of $h$. 

\vspace{0.4cm} 
\noindent 
A large $R$ expansion can be obtained by using the formula 
\beq 
\delta\left\langle X\right\rangle =-\frac1{4\pi}\int\left\langle 
\theta(x)X\right\rangle e^{\chi(x)}\delta\chi(x)d^{2}x\label{genvar} 
\eeq 
which gives the variation of $\left\langle X \right\rangle$ in terms of  
insertions of $\theta (x)$. 
By applying  this formula to $\theta (0)$  
one gets 
\beq 
\left\langle \theta(0)\right\rangle _{\textrm{sphere}}\sim 4  
\pi \mathcal{E}_{\textrm{vac}}+\frac{b_{1}}{R^{4}}+\frac{2b_{2}}{R^{6}}+\ldots\label{tcorr} 
\eeq 
where  $A$ is defined in terms of vacuum energy in flat  
space 
\beq 
\mathcal{E}_{\textrm{vac}}=-A\lambda^{1/(1-\Delta_\sigma)}\label{Evac} 
\eeq  
and the coefficients   
 $b_i $ can be expressed in terms of 
integrals of higher flat correlations functions of $\theta$.  
 
The derivative  of the partition function with respect to R 
is given by 
\beq 
\frac{d\log Z_{\lambda}(R)}{dR^{2}}=-\left\langle \theta\right\rangle 
\label{theta} 
\eeq 
and, combining (\ref{tcorr}) and (\ref{theta}) it follows that 
\beq 
\log\frac{Z_{\lambda}(R)}{Z_{0}}\sim-4\pi R^{2}\mathcal{E}_{\textrm{vac}} 
+\log(z_{\infty})+\frac{b_{1}}{R^{2}}+\frac{b_{2}}{R^{4}}+\ldots\label{ZIR} 
\eeq 
It is better to express everything in terms of $h$ to get 
\beq 
\log z(h)=\pi Ah^{1/(1-\Delta_\sigma)}+\log(2^{c/3}z_{\infty})-\frac c{6-6\Delta_\sigma}\log 
h+a_{1}h^{-1/(1-\Delta_\sigma)}+\ldots\label{zIR}%
\eeq 
where $z_{\infty} = \lambda^{c/(6-6\Delta_\sigma )}Z_{\infty}$ and $a_1 = 4b_1 
\lambda^{(1/(1-\Delta_\sigma)} \ldots$  
 
\subsection{Vacuum expectation values} 
\label{3.2} 
We will be interested also in one-point functions of relevant operators (VEV), 
which in flat space give essential informations about short distance expansion 
of correlation functions. 

By expressing $\lambda$ in terms of $h$ and defining 
\beq   
G_\Phi(h) \equiv(2 R)^{2 \Delta_\sigma } \left\langle \Phi (0)\right\rangle_{\lambda}
\eeq 
one can show that
\beq 
G_\Phi(h)=\sum _{n=1}^{\infty} g_n (-h)^n 
\eeq 
\noindent
where 
\beq
g_n = \frac{1}{ (2\pi) ^n n! } \int \left\langle \Phi(0) \sigma (x_1 )......\sigma (x_n) 
\right\rangle _{c} \prod_{i=1}^{n} 
\frac{d^2 x_i}{(1+x_i \bar{x}_i)^{2-2 \Delta}}
\eeq
and the subfix $c$ means connected with respect to $\Phi(0)$.	

\vspace{2cm}  

\noindent
Using (\ref{genvar}) it is also possible to write  a large $R$ expansion 
of the form 
\beq 
G_\Phi(h) =  
A_\Phi h^{\frac{\Delta_\Phi }{1-\Delta_\sigma }} + a_1 h^{-\frac{\Delta_\Phi
-1}{\Delta_\sigma -1}} + \ldots 
\eeq 
\noindent 
We recall that, in our normalizations, the VEV of a primary field on the plane is given by
\beq
\langle \Phi \rangle_\lambda = A_\Phi \left( \frac{\lambda}{2 \pi} 
\right)^{\frac{\Delta_\Phi }{1-\Delta_\sigma }}.
\eeq
There exists  a  simple relation 
between the VEV of the perturbing operator and the derivative of the 
partition function 
\beq 
G_{\sigma} (h) = -2 \frac{z'(h)}{z(h)} = \frac{d}{dh} \log z(h) 
\eeq 
and if $z(h)$ is an entire function the same is true 
for $z'(h)$ (the unnormalized VEV of the perturbing operator). It follows that, the asymptotic expansion for $G_{\sigma} (h)$ is given by
\beq
G_{\sigma} (h)=-\frac{2 \pi A}{1-\Delta_\sigma} h^{\frac{\Delta_\sigma}{(1-\Delta_\sigma)}}+\frac c{3-3\Delta_\sigma} h^{-1}+\ldots \, .
\eeq
 As a result, in the $h \to \infty$ limit, the usual relation between $A$ and the amplitude of the perturbing operator holds
\beq
A_\sigma= \frac{A}{1-\Delta_\sigma} .
\label{usua}
\eeq
\section{Numerical Results} 
\subsection{VEVs on the plane} 
The TCS approach enables us to study numerically the  
behaviour of both the VEV's and the partition function of the model.  
The aim of this section is to give an estimate of the VEV's of the  
primary operators of the Ising model, i.e. magnetization and energy,  
in the limit of infinite plane by means of the asymptotic formul\ae$~$of  
sect.\ref{3.2}. Since the model is defined on the sphere, in the large $h$  
limit we shall be able to recover the amplitudes $A_\Phi$, with $\Phi \equiv \sigma, \epsilon$, obtained on the plane, whose value is exactly known \cite{russiVEV}.\\ 
Our strategy proceeded as follows: first, we fit the data obtained from TCS technique  
by means of the expansion of sect.\ref{3.2} at finite values of the truncation level $N$ (we considered $N=10,$ $11,$ $12,$ $13,$ $14$) in an interval of the variable $h$ ranging from $150$ to $200$ (this choice is motivated by the requirement to be in an asymptotic region where truncation artifacts are not present); second, we perform  an extrapolation of the VEV's $A_\phi (N) $ to the limit $N \to \infty$ by means of the following (conjectured, see \cite{Fonseca:2001dc}) law 
\beq 
A_\phi (N) \ = \ A_\phi (\infty) + A_\phi^1 N^{-x} + \dots 
\eeq 
where the constant $A_\phi (\infty)$ is the extrapolated value of the amplitude $A_\Phi$. In order to get rid of possible systematic errors, the fitting procedure we used is the same as \cite{gr1}.
In this way, we obtained the estimates for the amplitudes\footnote{Their actual value is written using the standard normalization, see e.g. \cite{russiVEV}, where the factor $2 \pi$ in (\ref{cpert}) is absent. It is simply obtained by replacing $A_\Phi \to A_\Phi (2 \pi)^{\frac{\Delta_\Phi}{\Delta_\sigma-1}} $.} $A_\sigma$ and $A_\epsilon$
\beq 
A_\sigma=1.27759(6), \ \ \ \ \ \ \ \ \ \ \ A_\epsilon=2.004(8) 
\eeq 
which are in perfect agreement with both their theoretical values and the existing numerical estimates. 
\begin{table}
\begin{center}
\begin{tabular}{ll}
\hline
\hline
&\\[-5mm]
Asymptotic &$ N=10$\\
\hline
\hline
&\\[-5mm]
$3.01717$&$3.06882$\\
$8.01896$&$8.03176$\\
$12.8052$&$12.8043$\\
$17.4721$&$17.4603$\\
$22.0563$&$22.0296$\\
$26.5773$&$26.5378$\\
$31.0474$&$30.9807$\\
$ 35.4748$&$35.3684$\\
$39.8655$&$39.6865$\\
$44.2242$&$43.9406$\\
\hline
\hline
\end{tabular}
\hspace{-3mm}
\begin{tabular}{l}
\hline
\hline
\\[-5mm]
$N=11$\\
\hline
\hline
\\[-5mm]
$3.06886$\\
$8.03233$\\
$12.8058$\\
$17.4631$\\
$22.0334$\\
$ 26.5457$\\
$ 31.0007$\\
$35.4153$\\
$39.7763$\\
$44.0895$\\
\hline
\hline
\end{tabular}
\hspace{-3mm}
\begin{tabular}{l}
\hline
\hline
\\[-5mm]
$N=12$\\
\hline
\hline
\\[-5mm]
$3.06888$\\
$8.03236$\\
$12.8064$\\
$17.4643$\\
$22.0353$\\
$ 26.5485$\\
$31.0053$\\
$35.4231$\\
$39.7903$\\
$44.1134$\\
\hline
\hline
\end{tabular}
\end{center}
\label{tab1}
\caption{The comparison between asymptotic formula (\ref{zr}) and TCS approach is shown at different values of the truncation level $N$. The agreement improves for higher values of $h$.}
\end{table}
\begin{figure}
\begin{center}
\includegraphics[width=11cm]{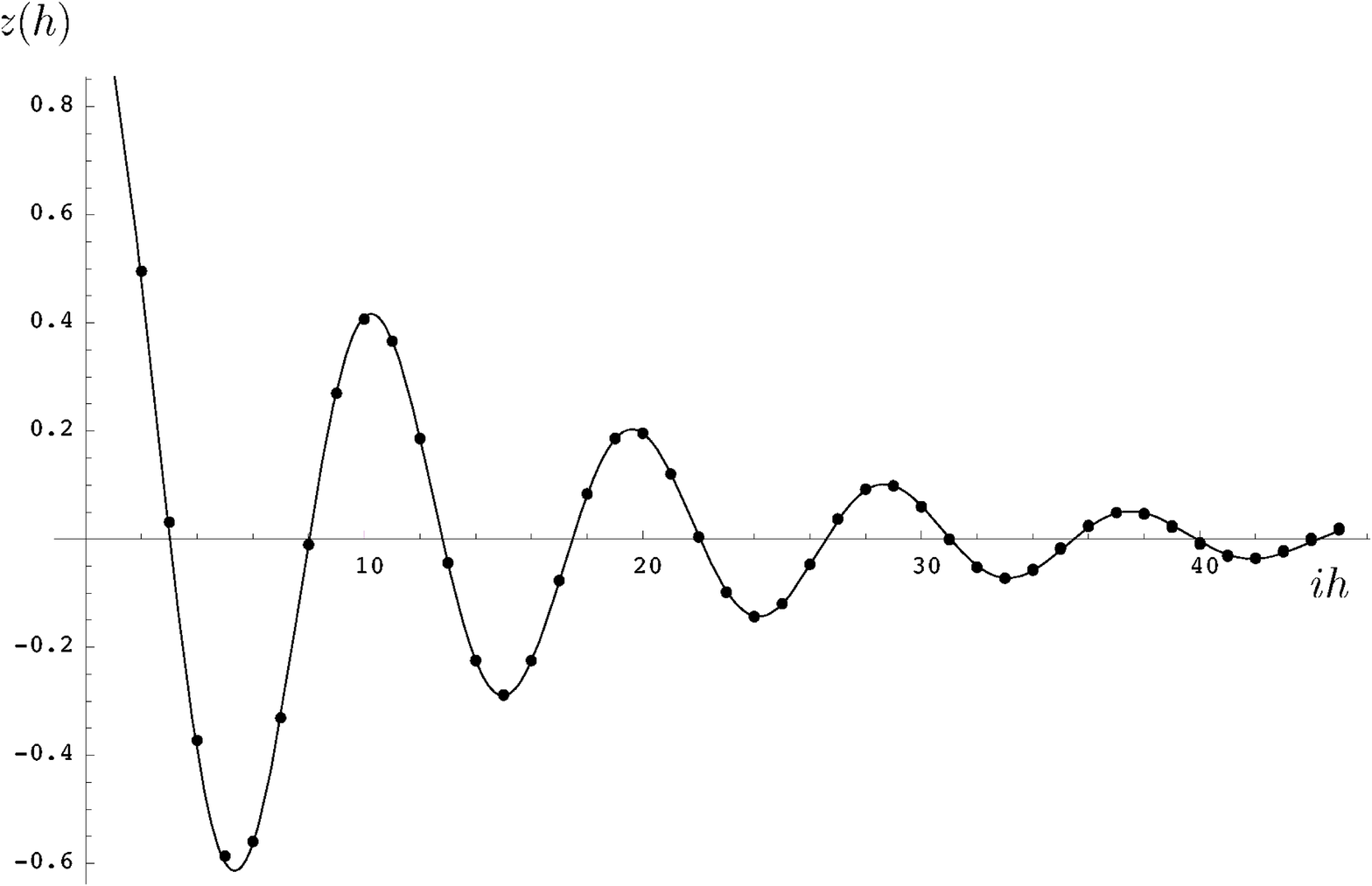}
\end{center}
\label{fig1}
\caption{Behaviour of the partition function $z(i h)$. The solid line represents the asymptotic behaviour and dots show TCS approach results.}
\end{figure}
\subsection{Analyticity properties of the partition function} 
Let us consider the case of pure imaginary values of the coupling $h$. One can see that the asymptotic behaviour of the partition function changes dramatically, becoming oscillatory and showing a well defined pattern of zeros. Such zeros are approximatively located at (when the leading order in the asymptotic expansion is considered, see \cite{zamo}) 
\eqa
-i h_n^{(a)} & = & \left[ A \pi \sin \left( \frac{\pi }{2 -2 d} \right) \right]^{d-1} \left( \frac{\pi \; c}{12 -12 d} +\frac{\pi}{2}+n \pi 
\right)^{1-d} = \nonu \\
& = & \left[ A  \sin \left( \frac{8 \pi }{15} \right) \right]^{-15/16} \left( n + \frac{49}{90}  \right)^{15/16}, \ \ \ \ \ n \ge 0
\label{zr} 
\ena
where for the Ising model we have $c=1/2$, $d=1/16$, $A=0.168564 \dots$ (we used the same notation as \cite{zamo}).  
Hence, one could ask whether the hypothesis of \cite{zamo} to consider the partition function as an entire function is compatible with numerical data also in the present case. Taking advantage of the truncated conformal space approach, we were able to compute numerically the partition function for the following values of the truncation level $N=10$, $11$, $12$. Figure \ref{fig1} shows such numerical results together with the plot of the asymptotic expansion (truncated at the leading order). Furthermore, one can compare the asymptotic location of zeros with the corresponding numerical estimates coming from the TCSA. The result of such comparison is shown in table \ref{tab1}. 
It is interesting to check the validity of (\ref{zr}) against the exact sum rule \cite{zamo} 
\beq 
\sum_{n=0}^\infty \; \frac{1}{h_n^2} = \frac{1}{7}= 0.142857 \dots 
\eeq       
Plugging (\ref{zr}) in the previous expression, we obtain 
\beq 
\sum_{n=0}^\infty \; \frac{1}{(h_n^{(a)})^2} = 0.146549 \dots 
\eeq 
which is under the 3\% of accuracy with respect to the exact result. \\ 
The previous findings strongly suggest that the partition function can  
be considered as an entire function of the coupling $h$.   \\
As a final remark, we could ignore the exact knowledge of $A$ and try estimate it taking advantage of both the sum rules and formula (\ref{zr}). The numerical result
\eqa
A{\textrm{\tiny asy}}= \frac{1}{\sin \frac{8}{15} \pi} \, \left ( 7 ~ \zeta (15/8,49/90) \right)^{-8/15} = 0.1663 \dots
\nonu
\ena
is remarkably near to the exact one $A=0.168564 \dots$.
\section{Conclusions}
Our results show that the partition function of the Ising model on the sphere may have interesting analytical properties in the
coupling $h$. \\
The present case, together with the Yang-Lee model considered by Al. Zamolodchikov in \cite{zamo}, seems to suggest
the conjecture that \textsl{all} the perturbed Rational Conformal Field Theories (when the perturbation is relevant)
share similar analytic properties in their observables.
In particular a proof of the convergence of the series defining the partition function should be welcome.
An attempt in this direction could be tried along the same lines of the proof of
convergence of strongly relevant ($ 2 \Delta < 1$) perturbations of RCFTs in finite volume \cite{flume}. \\
It would be also interesting to establish if there is a relation beetween the integrability of these models
on the plane and their analytical properties on the sphere.
In this perspective, some clarification
could come from the study of a non-integrable perturbation\footnote{Costantinescu and Flume \cite{flume}
shown that Conformal Perturbation Theory is convergent in finite volume
(which is similar to the present case) weather the theory is integrable or not.
On the other hand, McCoy \cite{McCoy:2000gw} pointed out that the results
of Orrick et al \cite{orrick} seem to suggest that
the convergence of Conformal Perturbation Theory fails if non-integrable perturbations are considered.
A careful analysis of this apparent contraddiction should be interesting, however it is beyond the purpouses of this letter.}
of a given minimal model (e.g., the most relevant magnetic perturbation of the Tricritical Ising model).
Finally, we would like to stress that the VEV of primary (relevant)
operators in the limit of infinite plane ($R \to \infty$) can be extracted with good precision using TCS approach.
\vskip0.4cm
\noindent
\section*{Acknowledgements}
We would like to thank Al. Zamolodchikov for a useful discussion at the beginning of this work,
A. Delfino and M. Caselle for a careful reading of the paper and F. Gliozzi for useful discussions. The work of P.G. is supported by the COFIN ``Teoria dei Campi, Meccanica Statistica e Sistemi Elettronici''.
 

\begin{thebibliography}{99} 
\bibitem{birrel}
N. B. Birrell and P.C.W. Davies, Quantum fields in curved space, Cambridge University Press (1984).
\bibitem{zamo}  
A. Zamolodchikov, hep-th/0109078. 
\bibitem{eguchi} 
T. Eguchi and H. Ooguri,  Nucl.\ Phys.\ B {\bf 282}  (1987) 308.
\bibitem{Fonseca:2001dc}
P.~Fonseca and A.~Zamolodchikov,
arXiv:hep-th/0112167.
\bibitem{cardy}
J. Cardy and L. Peshel,  Nucl. \ Phys. \ B {\bf 300} (1988) 377.
\bibitem{russiVEV}
S.~Lukyanov and A.~B.~Zamolodchikov,
Nucl.\ Phys.\ B {\bf 493} (1997) 571.
[arXiv:hep-th/9611238];
V.~Fateev, S.~Lukyanov, A.~B.~Zamolodchikov and A.~B.~Zamolodchikov,
Nucl.\ Phys.\ B {\bf 516} (1998) 652.
[arXiv:hep-th/9709034].
\bibitem{gr1} M.~Caselle and M.~Hasenbusch,
Nucl.\ Phys.\ B {\bf 639} (2002) 549.
[arXiv:hep-th/0204088];
M.~Caselle and M.~Hasenbusch,
Nucl.\ Phys.\ B {\bf 579} (2000) 667.
[arXiv:hep-th/9911216];
P.~Grinza and A.~Rago,
arXiv:cond-mat/0210046;
P.~Grinza and A.~Rago,
Nucl.\ Phys.\ B {\bf 651} (2003) 387.
[arXiv:hep-th/0208016].
%
\bibitem{McCoy:2000gw}
B.~M.~McCoy,
arXiv:cond-mat/0012193.
\bibitem{orrick}
W.P. Orrick, B. Nickel, A.J. Guttmann and J.H.H. Perk, J. Stat. Phys. 102 (2001) 795.
\bibitem{flume}
F. Constantinescu and R. Flume, Physics Letters B {\bf 326} (1994) 101.
\end{thebibliography}
\end{document}